\documentstyle[12pt,epsf,epsfig]{article}
\newcommand{\k}{{\sf k}}                   
\newcommand{\pp}{{\sf p}}                  
\newcommand{\uq}{{\sf u}}                  
\newcommand{\dq}{{\sf d}}                  
\newcommand{\sq}{{\sf s}}                  
\newcommand{\cq}{{\sf c}}                  
\newcommand{\bq}{{\sf b}}                  
\newcommand{\qq}{{\sf q}}                  
\def\la{{$\Lambda$}}
\def\al{{$\overline{\Lambda}$}}
\def\lal{{$\Lambda / \overline{\Lambda}$}}

\begin{document}

\date{\today}

\vspace{1cm}
\title{\Large \bf $\Lambda$ and $\overline{\Lambda}$ Polarization
in Lepton Induced Processes}

\vspace{1cm}
\author{\large A.~Kotzinian,$^\ast$ A.~Bravar, and D.~von~Harrach \\
\\
Institut f\"ur Kernphysik, Universit\"at Mainz,
D-55099 Mainz, Germany}

\vspace{1cm}
\maketitle

\vspace{2cm}
\begin{abstract}
The study of the longitudinal polarization of \la~and \al~hyperons produced
in polarized deep inelastic scattering,
neutrino scattering, and in $Z^0$ decays allows one to access the spin
dynamics of the quark fragmentation process.
Different phenomenological spin transfer mechanisms are considered
and predictions for the \la~and \al~longitudinal polarization 
in various processes using unpolarized and polarized targets are made.
Current and future semi-inclusive deep inelastic scattering experiments
will soon provide accurate enough data to study these phenomena
and distinguish between various models for the spin transfer mechanisms.
\\
\\
\end{abstract}

\vspace*{\fill}
\begin{center}
\end{center}

\vspace*{10mm}

\noindent $^\ast$ On leave from Yerevan Physics Institute,
375036 Yerevan, Armenia, \\and JINR, 141980 Dubna, Russia 

\newpage
\section{Introduction}

Sensitive tests of strong interaction dynamic models are provided
by polarization measurements.
The largest amount of discussions have been stimulated by the polarized
deep inelastic lepton nucleon scattering (DIS) measurements~\cite{DIS}. 
These results suggest that the angular momentum of the nucleon is not 
distributed among its parton constituents in the way expected in 
na{\"{\i}}ve quark models.
New information on the non-perturbative dynamics of strong interactions 
can be obtained by investigating semi-inclusive deep inelastic processes
which, in addition to the nucleon parton distribution functions,
depend also on fragmentation functions.

To investigate the spin transfer phenomenon in quark fragmentation one needs
a {\it source} of polarized quarks. The simplest processes involving
polarized quark fragmentation are the $e^+e^-$ annihilation, the DIS of
polarized charged leptons off unpolarized and polarized targets,
and neutrino (anti-neutrino) DIS.  
The self-analyzing decay properties of the \lal~hyperon make this
particle particularly interesting for spin physics.
A first theoretical study of the \la~polarization in hard processes  
\begin{equation}
l + N  \rightarrow l^\prime + \Lambda / \overline{\Lambda} + X
\label{eq:ln}
\end{equation}
and
\begin{equation}
e^+ + e^- \rightarrow \Lambda / \overline{\Lambda} +X
\label{eq:ee}
\end{equation}
to investigate the longitudinal spin transfer from polarized quarks
(di-quarks) to \lal's was made by Bigi~\cite{bi}.
The idea of using \lal's as a quark transverse-spin polarimeter in
reaction~(\ref{eq:ln}) was originally proposed by 
Baldracchini~{\it et al.}~\cite{bgrs},
and later rediscovered by Artru and Mekhfi~\cite{artm}.
The \lal~polarization in reactions (\ref{eq:ln}) and (\ref{eq:ee})
has been discussed in several recent works~\cite{gh}--\cite{Jaf96},
which show a considerable interest for this problem.
Some \lal~polarization data already exist for the reaction (\ref{eq:ln})
from neutrino (anti-neutrino) beam experiments~\cite{neut}
and for the reaction~(\ref{eq:ee}) at the $Z^0$ pole~\cite{aleph}.
New high statistics data are expected soon from several 
experiments~\cite{hermes,e665,nomad,comp}.
 
The longitudinal polarization transfer mechanism from a polarized lepton to
the final hadron in reaction~(\ref{eq:ln}) is based on the idea~\cite{bi},
that the exchanged polarized virtual boson will strike preferentially
one quark polarization state inside the target nucleon,
and that the fragment left behind will contain some 
memory of the angular momentum removed from the target nucleon,
thus resulting in a non-trivial longitudinal polarization of $\Lambda$
hyperons produced in the target fragmentation region ($x_F < 0$,
$x_F$ is the Feynman $x$)~\cite{EKK96}.
The fragmenting struck quark in turn can transfer its polarization to
a \lal~hyperon produced in the current fragmentation region ($x_F > 0$).
In both cases the underlying dynamics of the hyperon production and
polarization cannot be described by perturbative QCD and some
phenomenological models have to be considered.

A phenomenological study of the \la~and \al~longitudinal polarization
in the reaction
\mbox{$\mu^+ + N \rightarrow \mu^{+\prime} + \Lambda / \overline{\Lambda} + X$}
has been already presented by us in the {\it COMPASS} proposal~\cite{comp}.
Here we present a more detailed and complete study of the
\lal~polarization in DIS of charged leptons and neutrinos in the current
fragmentation region and in $e^+e^-$ annihilation at the $Z^0$ pole.
The measurement of the \lal~polarization in these processes can help us to
distinguish between different mechanisms of the spin transfer 
in the quark fragmentation. 
In Section~2 we describe some models for the spin transfer mechanism
that we consider in our studies.
In Sections~3 and 4 we present predictions for
the \lal~polarization in electro-production with polarized leptons on unpolarized and polarized targets,
in Section~5 for neutrino and anti-neutrino scattering,
and in Section~6 at the $Z^0$ pole.
Section~7 contains a discussions of the results presented in this work. 

\section{Models for spin transfer in quark fragmentation}

The quark fragmentation functions as well as the parton distribution functions
of the nucleon are well defined objects in quantum field theory.
The spin, twist, and chirality structure of the quark fragmentation functions,
integrated over the transverse momentum, are discussed
and classified in~\cite{jj}.
The leading twist unpolarized ($D_q^\Lambda (z)$)
and polarized ($\Delta D_q^\Lambda (z)$) quark fragmentation
functions to a \la~hyperon are defined as:
\begin{eqnarray}
D_q^{\Lambda}(z)&=& D_q^{+ \;\Lambda}(z)+ D_q^{- \;\Lambda}(z) \\
\Delta D_q^{\Lambda}(z)&=& D_q^{+\;\Lambda}(z)- D_q^{-\;\Lambda}(z)
\end{eqnarray}
where  $D_q^{+ \;\Lambda} (z)$ ($D_q^{- \;\Lambda} (z)$) is the spin
dependent quark fragmentation functions for the \la~spin parallel (anti-parallel) to that of the initial quark $q$, and $z$
is the quark energy fraction carried by the $\Lambda$ hyperon.

We will parametrize the polarized quark fragmentation functions as 
\begin{equation}
\Delta D_q^\Lambda (z) = C_q^\Lambda (z) \cdot D_q^\Lambda (z)
\end{equation}
where $C_q^\Lambda (z)$ are the spin transfer coefficients.
Since much is still unknown on polarized fragmentation functions,
we do not consider explicitly their $Q^2$ evolution in this work
(see for instance Ref.~\cite{Rav96}).
In the literature there exists some models~\cite{bfm80,nh95}
for the spin dependent fragmentation functions in which a
$z$ dependence of the spin transfer coefficients can be found.
For example, in the jet fragmentation model of~\cite{bfm80},
$C_q^\Lambda (z) \sim z$ at small $z$
and $C_q^\Lambda (z) \rightarrow 1$ at $z \rightarrow 1$.
In the covariant quark -- di-quark model of~\cite{nh95}
$C_u^\Lambda (z) \sim z$ at small $z$,
whereas $C_s^\Lambda (z) \sim const$.
We will not present here predictions for the \lal~polarization
obtained with these models, since they contain many free
parameters which are not well tuned with existing data.

To get quantitative predictions for the \lal~polarization in processes 
(\ref{eq:ln}) and (\ref{eq:ee}) we used a phenomenological approach
similar to that of Ref.~\cite{gh}.
We consider two different descriptions of the spin transfer mechanism
in the quark fragmentation to a \lal~hyperon.
The first one is based on the non-relativistic quark model SU(6)
wave functions, where the \la~spin is carried only by its constituent
\sq~quark.
Therefore, the polarization of directly produced \la's is determined
by that of the \sq~quark only, while \la's coming from decays of heavier
hyperons inherit a fraction of the parent's polarization, which
might originate also from other quark flavors
(namely \uq~and \dq).
In this scheme the spin transfer is discussed in terms of
{\it constituent quarks}.
Table~\ref{tab:bgh} shows the spin transfer coefficients
$C_q^\Lambda$ for this case~\cite{bi,gh}.
As discussed in Section 6 and shown in~\cite{aleph}, this model
reproduces fairly well the \lal~longitudinal polarization measured
at the $Z^0$ pole and at large $z$. However, the interpretation of these
data is not unique.
A particular case is given by a simpler assumption that the \la~hyperon
gets its polarization from \sq~quarks only.
In the following we will refer to the former description as $BGH$
({\it for} Bigi, Gustafson, and H\"akkinen)
and the latter as $NQM$ ({\it for} na\"{\i}ve quark model).

\begin{table}
\begin{center}
\begin{tabular}{|c||c|c|c|c|}
\hline
\la's parent & $C_u^\Lambda$ & $C_d^\Lambda$ & $C_s^\Lambda$ & 
$C_{\bar q}^\Lambda$\\ \hline \hline
Quark          & $0$    & $0$    & $+1$   & $0$ \\ \hline 
$\Sigma^0$     & $-2/9$ & $-2/9$ & $+1/9$ & $0$ \\ \hline 
$\Sigma(1385)$ & $+5/9$ & $+5/9$ & $+5/9$ & $0$ \\ \hline
$\Xi$          & $-0.3$ & $-0.3$ & $+0.6$ & $0$ \\ \hline
\end{tabular}
\end{center}
\caption{Spin transfer coefficients according to non-relativistic SU(6)
quark model.}
\label{tab:bgh}
\end{table}

\begin{table}
\begin{center}
\begin{tabular}{|c||c|c|c|c|}
\hline
  & $C_u^\Lambda$ & $C_d^\Lambda$ & $C_s^\Lambda$ & $C_{\bar q}^\Lambda$ \\ 
\hline \hline
{\it BJ-I}    & $-0.20$  & $-0.20$  & $+0.60$ & $0.0$  \\ \hline
{\it BJ-II}   & $-0.14$  & $-0.14$  & $+0.66$ & $-0.06$  \\ \hline
\end{tabular}
\end{center}
\caption{Spin transfer coefficients according to the Burkardt-Jaffe 
$g_1^\Lambda$ sum rule.}
\label{tab:bjsr}
\end{table}

The second approach is based on the $g_1^\Lambda$ {\it sum rule} for the
first moment of the polarized quark distribution functions in
a polarized \la~hyperon, which was derived by Burkardt and Jaffe~\cite{BuJ93}
in the same fashion as for the proton one ($g_1^{\pp}$).
We assume that the spin transfer from a polarized quark \qq~to
a \la~is proportional to the \la~spin carried by that flavor,
{\it i.e.} to $g_1^\Lambda$.
Table~\ref{tab:bjsr} contains the spin transfer coefficients $C_q^\Lambda$,
which were evaluated using the experimental values for $g_1^{\pp}$.
Two cases are considered~\cite{Jaf96}:
in the first one only valence quarks are polarized;
in the second case also sea quarks and anti-quarks
contribute to the \la~spin.
In the following we will refer to the first one as {\it BJ-I} and the
second one as {\it BJ-II}.
In this description, \la's~originating from strong decays of hyperon
resonances are absorbed in the \la~fragmentation function.
A similar description for $\Sigma^0$'s and cascades is not yet available;
therefore \la's originating from decays of these hyperons
have to be excluded in this description.
As our calculations have shown,
the exclusion of these \la's has a small effect on the final
polarization result (contained to within a few~\%).

\begin{figure}
\vspace*{-10mm}
\begin{center}
\mbox{\epsfxsize=16cm\epsffile{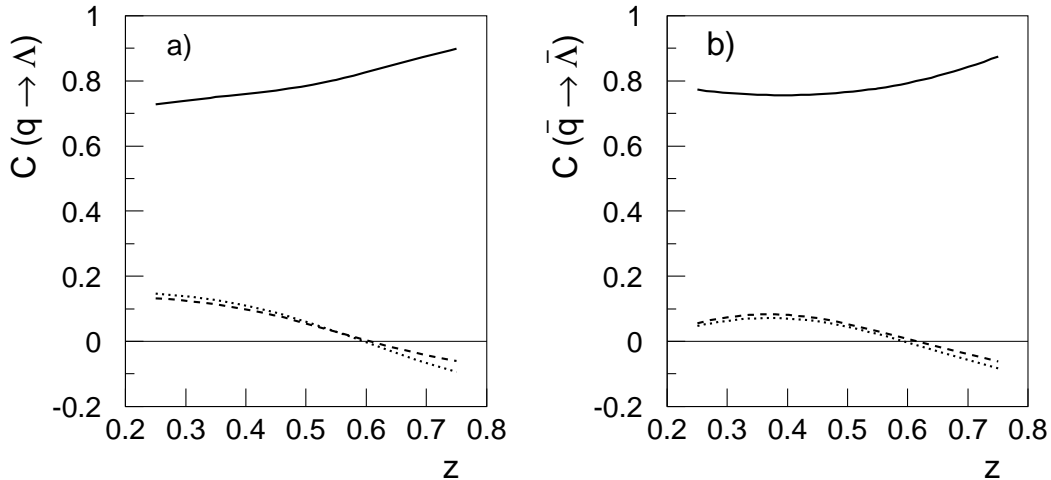}}
\end{center}
\vspace*{-5mm}
\caption{$z$ {\it dependence} of the spin transfer coefficients
$C_q^\Lambda$ in the {\it BGH} spin transfer mechanism.
a) $\Lambda$: solid line - \sq~quark, dashed - \uq, and
dotted - \dq;
b) ${\overline \Lambda}$: solid line - ${\sf \bar s}$~quark,
dashed  - ${\sf \bar u}$, dotted - ${\sf \bar d}$.}
\label{fig:zeddep}
\end{figure} 

In the $g_1^\Lambda$ {\it sum rule} scheme a negative spin transfer from
\uq~and \dq~quarks to a \la~hyperon is predicted.
A negative spin transfer from \uq~and \dq~quarks of $-0.09$ was also
predicted in~\cite{AR95} using an effective QCD Lagrangian,
and in the covariant quark -- di-quark model of ~\cite{nh95}. 
This effect can be understood qualitatively even if the spin
of the \la~is determined by its constituent \sq~quark only:
in some cases the fragmenting \uq~or \dq~quark will
become a sea quark of the constituent \sq~quark, and the spin
of the constituent \sq~quark will be anti-correlated to the spin of
the fragmenting quark~\cite{EKKS95,EKK96}.
Another possibility occurs when the \la~is produced as a second rank
particle in the fragmentation of a \uq~or \dq~quark.
If the first rank particle was a pseudoscalar strange meson,
then the spin of the ${\sf \bar s}$ quark has to be opposite to that
of the \uq~(\dq) quark,
and since the \sq${\sf \bar s}$ pair created from the vacuum in the string
breaking is assumed to be in a $^3P_0$ state~\cite{And79},
the \sq~quark is also oppositely polarized to the \uq~or \dq~quark.
This last mechanism of the spin transfer can be checked by measuring the 
\la~polarization for a sample of events containing fast $K$ mesons.

We implemented the spin transfer coefficients $C_q^\Lambda$ given
in Tables~\ref{tab:bgh} and~\ref{tab:bjsr} in appropriate
Monte Carlo event generators for different processes
on the basis of the program information on
the flavor of the fragmenting quark and 
the \la~production process (directly produced or originating from decays).
For the simulation of DIS events (charged leptons and neutrinos)
we used the {\tt LEPTO v.6.3 - JETSET v.7.4}~\cite{ing,sos} event generator,
and for the $e^+ e^-$ annihilation at the $Z^0$ pole the
{\tt PYTHIA v.5.7 - JETSET v.7.4}~\cite{sos} event generator.
With a suitable choice of input parameters these event generators
reproduce well the distributions of various measured physical observables
and the particle yields.
The quark hadronization is described by the LUND string fragmentation
model~\cite{And83}.
We used the LUND modified symmetric fragmentation function with
default parameter settings~\cite{sos}.
Different fragmentation schemes were also considered,
like the independent fragmentation options in {\tt JETSET}~\cite{sos}.
They lead to similar results and conclusions.
We set the strangeness suppression factor to 0.20 
in agreement with recent experimental data~\cite{ssup}.

In the {\it BGH} approach the spin transfer coefficients for individual
channels are $z$ independent.
However, the effective spin transfer coefficient for a given quark flavor,
obtained by summing over all \la~production channels
appears to have a $z$ dependence.
Thus using the $C_q^\Lambda$ from Table~\ref{tab:bgh} together with
appropriate weights for different \la~production channels, as obtained
from the event generators, we automatically introduce a
$z$ dependence in $C_q^{\Lambda}(z)$ (see Figure~\ref{fig:zeddep}). 
For the $g_1^\Lambda$ {\it sum rule} spin transfer mechanism
we make the simplest assumption,
in which the spin transfer coefficients are $z$ independent. 
If we choose a $z$ dependence for $C_q^\Lambda$ similar to the one
proposed in~\cite{bfm80}, we will obtain smaller (larger) values of the
\la~polarization at small (large) $z$.

\section{\la~and \al~polarization in charged lepton DIS off
an unpolarized target}
 
The complete twist-three level description of spin-$1/2$ baryons 
production in polarized DIS is given in~\cite{Mul96}.
Here we will consider this process at leading order 
integrated over the final hadron transverse momentum.
In this approximation the magnitude of the \la~longitudinal polarization
is given by the simple parton model expression~\cite{comp,EKK96}
\begin{equation}
P_{\Lambda} \, (x,y,z) = P_{\Lambda}^{\parallel} \, (x,y,z) =
\frac{\sum_q e_q^2 \;
          [P_B D(y) q(x) + P_T \Delta q(x) ]
          \; \Delta D_q^{\Lambda}(z)}
     {\sum_q e_q^2 \;
          [q(x) + P_B D(y) P_T \Delta q(x) ]
          \; D_q^{\Lambda}(z)},
\label{eq:lambdap}
\end{equation}
where $P_B$ and $P_T$ are the beam and target longitudinal
polarizations, $e_q$ is the quark charge,
$q(x)$ and $\Delta q(x)$ are the unpolarized 
and polarized quark distribution functions, and
$D_q^\Lambda (z)$ and $\Delta D_q^\Lambda (z)$ are the unpolarized
and polarized fragmentation functions.
\begin{equation}
D(y)=\frac{1-(1-y)^2}{1+(1-y)^2}
\label{eq:dy}
\end{equation}
is commonly referred to as the longitudinal depolarization factor
of the virtual photon with respect to the parent lepton,
where $y$ is the energy fraction of the
incident lepton carried by the virtual photon
\footnote{Here and in the following the sign of the \la~polarization is
given with respect to the direction of the momentum transfer ({\it i.e.}
along the axis of the exchanged virtual boson in DIS).}.

For scattering off an unpolarized target Eq.~(\ref{eq:lambdap}) reduces to
\begin{equation}
P_\Lambda \, (x,y,z) =P_B D(y)\;\frac{\sum_q e_q^2 \;
          q(x) \; \Delta D_q^{\Lambda}(z)}
     {\sum_q e_q^2 \;
          q(x)\; D_q^{\Lambda}(z)}.
\label{eq:lambdaup}
\end{equation}
This expression is intuitively easy to understand since the final quark polarization, $P_{q^\prime}$, in polarized lepton-unpolarized quark scattering 
is given by the QED expression
\begin{equation} 
P_{q^\prime}=P_B D(y).
\label{eq:pq}
\end{equation}

We implemented Eq.~(\ref{eq:lambdaup}) and the spin transfer
coefficients $C_q^\Lambda$ from Tables~\ref{tab:bgh} and~\ref{tab:bjsr}
into the {\tt LEPTO} code to predict the \lal~polarization
for different models of the spin transfer mechanism.
Our calculations have been performed in experimental conditions similar
to that of the proposed {\it COMPASS} experiment~\cite{comp}:
hard DIS ($Q^2 > 4$~GeV$^2$) of negatively polarized $\mu^+$'s
($P_\mu = -0.80$) at $E_\mu = 200~{\rm GeV}$
\footnote{The beam energy chosen in the {\it COMPASS}
proposal~\protect\cite{comp} is 100~GeV.
No big differences are expected for $E_\mu = 100~{\rm GeV}$
compared to $E_\mu = 200~{\rm GeV}$.}
off an unpolarized isoscalar target ($^6$LiD).
To select \la's produced in the current fragmentation region we require
$x_F > 0$ and $z > 0.2$
\footnote{A different selection of the current fragmentation region
was proposed by Berger~\cite{Ber87} - {\it Berger criterium}. Basically,
to each $W^2$ value it corresponds a range in $z$, $z_{min} < z < 1$,
where it should be possible to measure the fragmentation functions:
for instance for $W^2 > 55 \, (> 23)~{\rm GeV}^2$, $z > 0.1 \, (> 0.2)$.
In our kinematical conditions this criterium is satisfied 
automatically by all selected events.}.
Additionally, to enrich the sample with events with a large spin transfer 
in lepton -- quark scattering, we restrict the virtual photon energy
range to $0.5 <y <0.9$, which gives $\langle D(y) \rangle \sim 0.8$.
In these studies we used the recent $MRSA^\prime$ unpolarized parton
distribution functions~\cite{MRSA}.
In the selected kinematical region the yields of \la's and \al's are similar
as is their kinematical spectra.
Roughly 10~\% of all produced \la's and one third of \al's survive the
$x_F$ and $z$ cuts.
This sample represents about 1~\% of the total DIS cross section
($\sigma \sim 10~{\rm nb}$ in these conditions).

\begin{figure}
\vspace*{-10mm}
\begin{center}
\mbox{\epsfxsize=16cm\epsffile{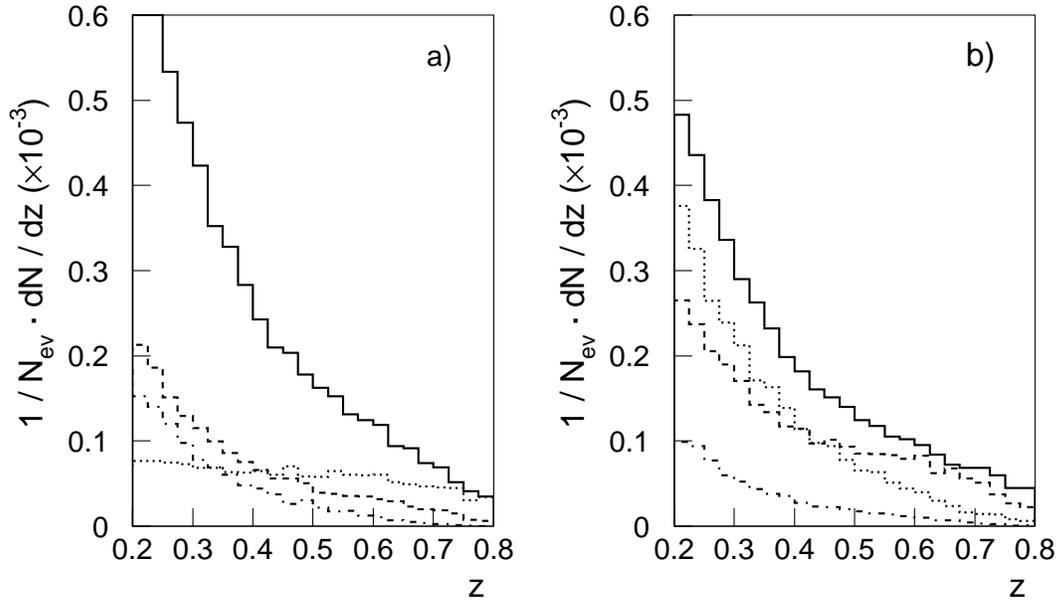}}
\end{center}
\vspace*{-5mm}
\caption{a) normalized $z$ distribution of \la's produced in $\mu^+$--DIS
originating from the fragmentation of different quark flavors:
solid line - \uq~quark, dashed - \dq, 
dotted - \sq, and dot-dashed - ${\sf \bar u}$;
b) normalized $z$ distribution of directly produced \la's (solid line),
\la's coming from $\Sigma(1385)$ resonances (dotted),
$\Sigma^0$ decays (dashed),
and cascades (dot-dashed).}
\label{fig:muzed}
\end{figure}

In Figure~\ref{fig:muzed}a we show the normalized $z$ distribution 
(to the total number of generated events)
of \la's created in the fragmentation of different quark
and anti-quark flavors as obtained from the {\tt LEPTO} code.
Figure~\ref{fig:muzed}b shows the $z$ distribution of directly produced \la's,
as well as \la's coming from $\Sigma^0$ decays, higher spin resonances
$\Sigma(1385)$, and cascades. Almost half of the total \la~sample is
produced directly.

\begin{figure}
\vspace*{-10mm}
\begin{center}
\mbox{\epsfxsize=16cm\epsffile{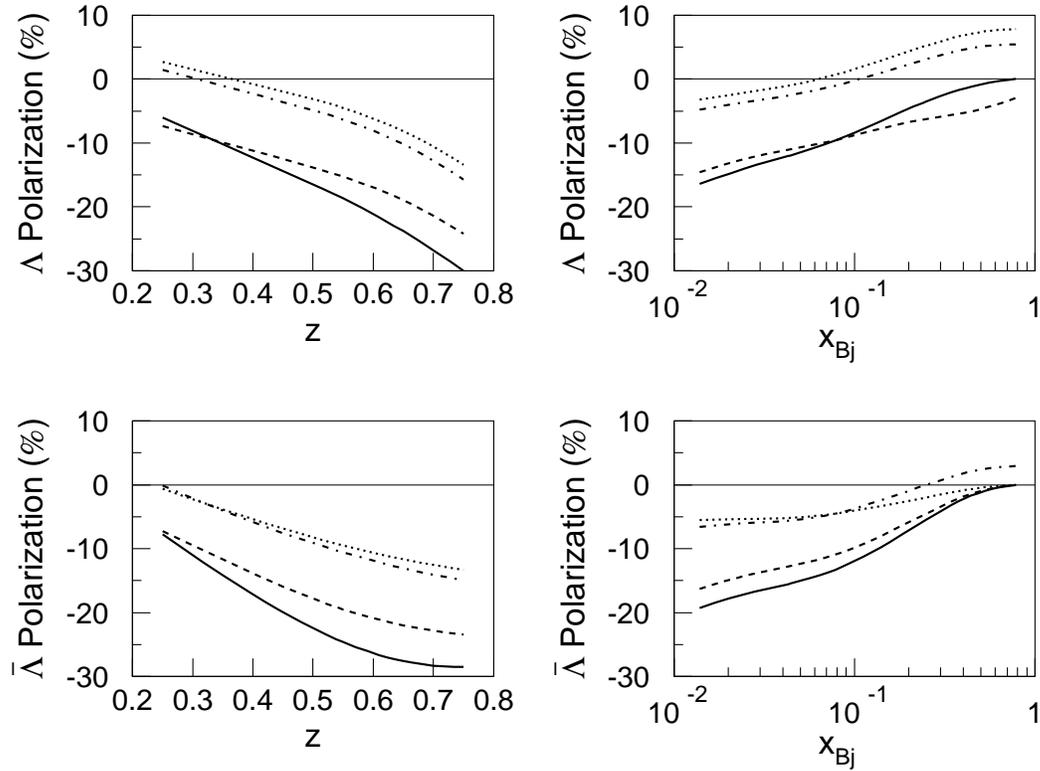}}
\end{center}
\vspace*{-7mm}
\caption{\la~and \al~longitudinal polarization in the current 
fragmentation region for DIS of polarized $\mu^+$'s on an unpolarized
target for different mechanisms of spin transfer:
solid line - $NQM$, dashed - $BGH$, dotted - {\it BJ-I}, and
dot-dashed - {\it BJ-II}.}
\label{fig:unpolarized}
\end{figure}

In Figure~\ref{fig:unpolarized} we present our results
for the \la~and \al~longitudinal polarization separately.
The differences in the \lal~polarizations are quite significant
between the two schemes for the spin transfer mechanism
(the {\it constituent quark} on one hand and
the $g_\Lambda^1$ {\it sum rule} on the other),
while they appear to be small (only a few~\%) within the same scheme.
Integrating in $z$ for $z > 0.2$ we expect a polarization of about
$-12~\% \, (-14~\%)$ for \la's (\al's) for the first scheme,
and $-2~\% \, (-5~\%)$ for the second one
\footnote{These results were obtained for a negative beam polarization of
$P_B = -0.80$. For different beam polarizations these results
can be rescaled accordingly.}.
No significant variations were observed for the \lal~polarization values,
when performing the same analysis for a proton or neutron target
in the same kinematical region.

Already at low values of $z$, the {\it constituent quark} scheme predicts
a sizeable negative \la~polarization, 
while the $g_\Lambda^1$ {\it sum rule} a slightly positive one.
At high $z$ both reach large negative values.
This behaviour of the \la~polarization is easily understood,
given that at low $z$ ($z < 0.5$) \la~production is dominated by
scattering off \uq~quarks (see Figure~\ref{fig:muzed}a), which in the
two schemes contributes to the \la~polarization with opposite signs
(see Table~\ref{tab:bgh} and Table~\ref{tab:bjsr}).
At high $z$ ($z > 0.5$) the relative contribution of \sq~quarks,
which contributes with the same sign but different magnitudes,
increases significantly, and eventually dominates at large $z$.
In the same way the $x$ dependence of the polarization can also be
understood: in the low $x$ region scattering off both \uq~and \sq~quarks
contribute to the \la~production and polarization,
while at high $x$ only \uq~quarks contribute.
A similar analysis applies to the \al~polarization results.

A high luminosity experiment, like {\it COMPASS}~\cite{comp}, 
might collect with these kinematical conditions a fully reconstructed
sample of \lal~in excess of $10^5$.
This will allow a precise determination of the \lal~polarization in
several bins over a wide $z$ range with a precision of a few~\% for each bin
and to distinguish between the two descriptions of the
spin transfer mechanism discussed above.

\begin{figure}
\vspace*{-10mm}
\begin{center}
\mbox{\epsfxsize=16cm\epsffile{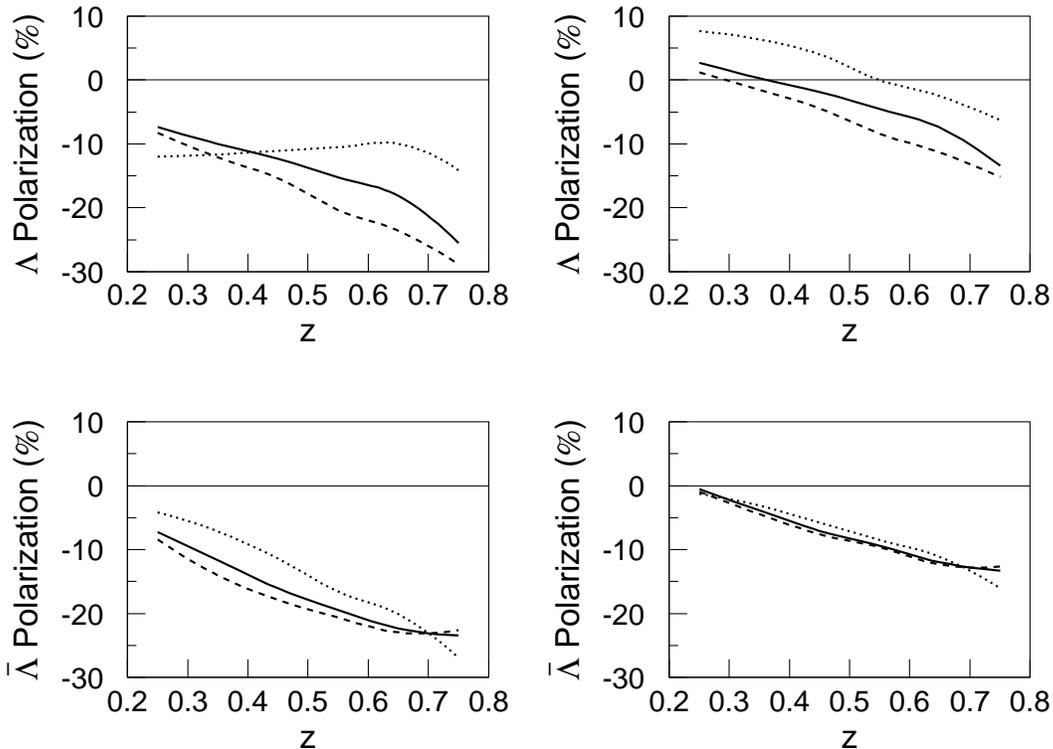}}
\end{center}
\vspace*{-7mm}
\caption{\la~and \al~longitudinal polarization for three different beam
energies using the {\it BGH} (left plots) and the {\it BJ-I} (right plots)
spin transfer mechanism:
solid line - $E_\mu = 200$~GeV, dashed - $E_\mu = 500$~GeV,
and dotted - $E_{el} = 30$~GeV.}
\label{fig:lampolen}
\end{figure} 

\begin{table}
\begin{center}
\begin{tabular}{|c||c|c|c|c|}
\hline
$E_{beam}$ (GeV) & $NQM$ & $BGH$ & $BJ-I$ & $BJ-II$ \\
\hline \hline
 30 -- $\Lambda$             &  $-5.8$ & $-12.0$ &  $4.9$ &  $2.4$  \\ \hline
 30 -- ${\overline \Lambda}$ & $-12.8$ & $-10.4$ & $-5.2$ & $-4.3$  \\ \hline 
\hline
200 -- $\Lambda$             & $-12.4$ & $-11.5$ & $-1.1$ & $-2.6$  \\ \hline
200 -- ${\overline \Lambda}$ & $-15.8$ & $-13.1$ & $-5.1$ & $-5.4$  \\ \hline 
\hline
500 -- $\Lambda$             & $-15.9$ & $-14.1$ & $-3.4$ & $-4.8$  \\ \hline
500 -- ${\overline \Lambda}$ & $-17.8$ & $-14.7$ & $-5.6$ & $-6.3$  \\ \hline 
\hline
\end{tabular}
\end{center}
\caption{\lal~longitudinal polarization for different lepton beam energies
and an unpolarized deuterium target, $x_F > 0$ and $z > 0.2$.}
\label{tab:lampolen}
\end{table}

We performed similar calculations also for different beam energies,
like the {\it HERMES} experiment~\cite{hermes} with the 30~GeV polarized
electron beam,
and the {\it E665} experiment~\cite{e665} with the 500~GeV polarized muon beam.
Table~\ref{tab:lampolen} summarizes these results for $x_F > 0$ and $z > 0.2$.
In Figure~\ref{fig:lampolen} we compare the \lal~polarization predictions
for these three different beam energies using the {\it BGH} and the {\it BJ-I}
spin transfer mechanism.
We always assume a beam polarization $P_B = -0.80$, an isoscalar target,
$Q^2 > 4~{\rm GeV}^2$, and $0.5 < y < 0.9$.
The conclusions are similar to the ones above,
except that at each different beam energy a different $x$ interval
is covered (the higher the energy, the lower the accessible $x$).
In particular, at 30~GeV (and $Q^2 > 4~{\rm GeV}^2$) \la~production
is dominated by scattering off \uq~quarks even at large $z$,
and the \la~polarization varies weakly with $z$ in the whole $z$ interval
for the {\it constituent quark} spin transfer mechanism,
since the accessible $x$ range hardly extends into the low $x$
region, where \sq~quarks are abundant. 

\section{\lal~polarization in DIS off a polarized target}

For a polarized target and polarized lepton beam (see Eq.~\ref{eq:lambdap})
there are two {\it sources} for the fragmenting quarks polarization:
the spin transfer from the polarized lepton and
from the struck polarized quark in the target.
We studied the polarization difference between the \lal~polarization
for positive (parallel to the beam polarization) and negative
target polarization (anti-parallel to the beam polarization)
\begin{equation}
\Delta P_\Lambda \, = \,
P_{\Lambda} \, (+ P_T) \, - \,
P_{\Lambda} \, (- P_T) \,.
\label{eq:delpol}
\end{equation}
By reversing the target polarization, the fragmenting quark polarization,
$P_{q'}$, changes by
\begin{equation}
\Delta P_{q'} (x,y) = 2 \, P_q(x) \,
\frac{1- (P_B \, D(y))^2}{1-(P_B \, D(y) \, P_q (x))^2}
\label{eq:poldif}
\end{equation}
where
\begin{equation}
P_q(x) = P_T \frac{\Delta q(x)}{q(x)}
\end{equation}
is the polarization of the quark in the polarized nucleon.

\begin{figure}
\vspace*{-10mm}
\begin{center}
\mbox{\epsfxsize=16cm\epsffile{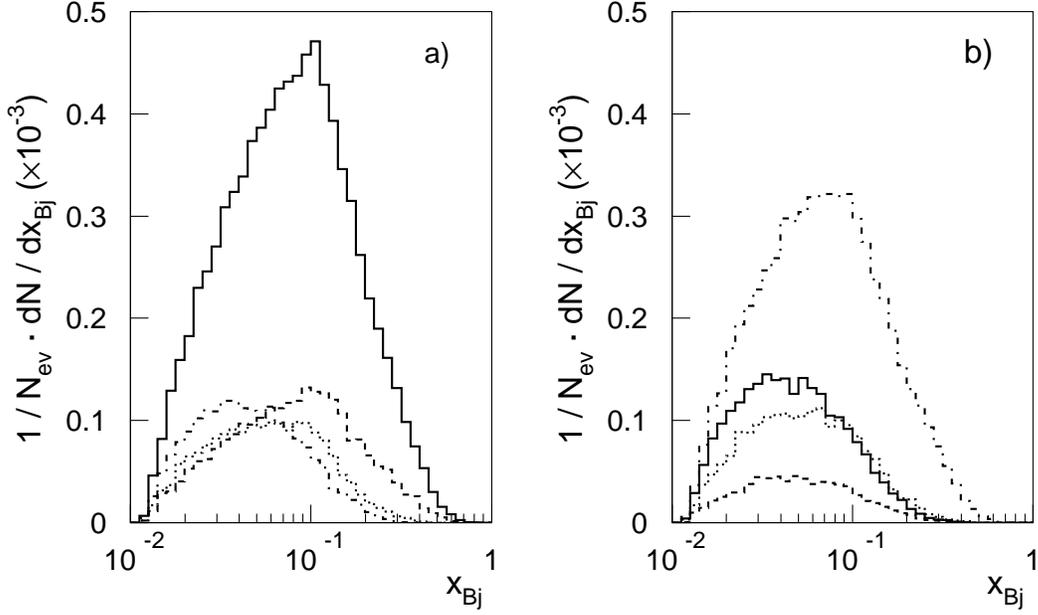}}
\end{center}
\vspace*{-5mm}
\caption{a) normalized $x$ distribution of \la's produced in
$\mu^+$--DIS from the fragmentation of different quark flavors:
solid line - \uq~quark, dashed - \dq, 
dotted - \sq, and dot-dashed - ${\sf \bar u}$;
b) same as (a) for \al's: solid line - ${\sf \bar u}$~quark,
dashed - ${\sf \bar d}$, dotted - ${\sf \bar s}$,
and dot-dashed - \uq~($x_F > 0$ and $z > 0.2$).}
\label{fig:muxbj}
\end{figure}

In our calculations we use a polarized proton target of $P_T = 0.80$
and a 200~GeV polarized $\mu^+$ beam of $P_B = -0.80$.
In most experiments complex target materials are used and the effective
nucleon polarization is significantly diluted
(for instance for a polarized $^6$LiD target
$\langle P_N \rangle \sim 25~\%$).
For such targets a smaller sensitivity
on the nucleon polarization is therefore expected
(roughly $3 \, \times$ smaller).
The covered kinematical range is similar to the one in the previous section.
To reduce the effects related to the beam polarization, 
we extend our analysis to the whole accessible $y$ range,
$0.1 \leq y \leq 0.9$,
to which corresponds a photon beam with smaller polarization
($\langle D(y) \rangle \sim 0.4$).
In $\Delta P_\Lambda$ ($\Delta P_{\overline \Lambda}$)
the beam polarization $(P_B \langle D(y) \rangle)$ enters quadratically,
and therefore affects little the final result 
(typically $(P_B \langle D(y) \rangle)^2 < 0.1$).
The total DIS cross section for this sample is about 35~nb.
Figure~\ref{fig:muxbj} shows the normalized $x$ distribution
of different quark and anti-qaurk flavors fragmenting to the
selected \la's and \al's.

\begin{figure}
\vspace*{-10mm}
\begin{center}
\mbox{\epsfxsize=16cm\epsffile{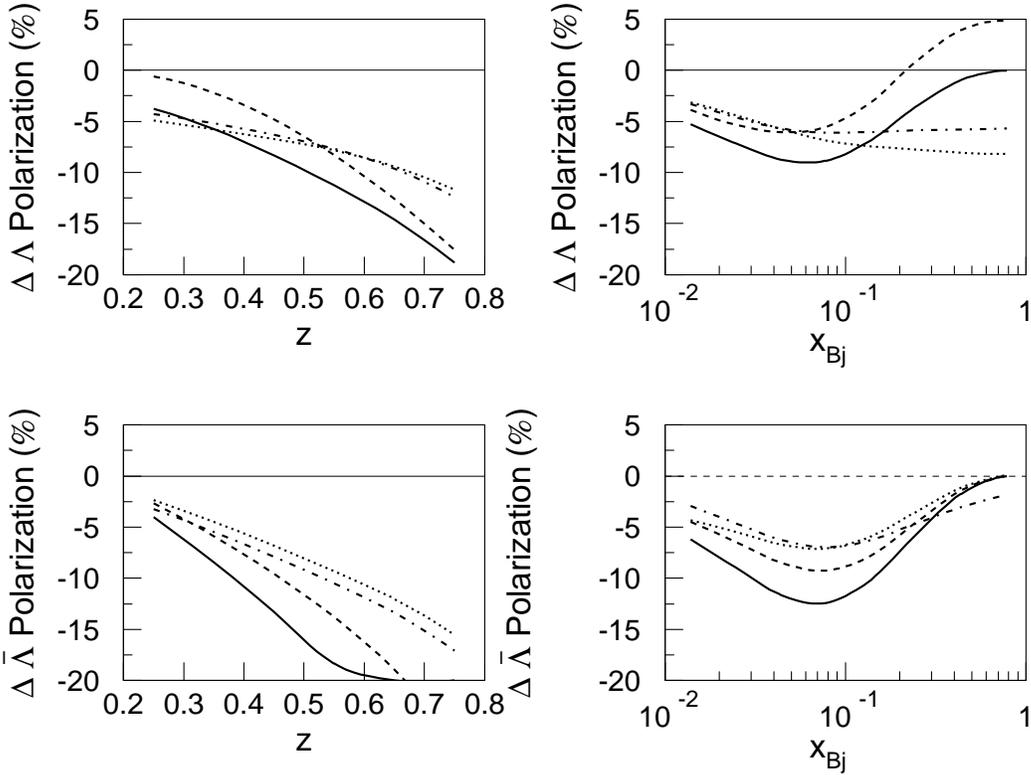}}
\end{center}
\vspace*{-7mm}
\caption{$\Delta P_\Lambda$ and $\Delta P_{\overline \Lambda}$
(Eq.~\protect\ref{eq:delpol}) for DIS
of polarized $\mu^+$'s off a polarized proton target:
solid line - $NQM$, dashed - $BGH$, dotted - {\it BJ-I}, and
dot-dashed - {\it BJ-II}.}
\label{fig:polarized}
\end{figure}

\begin{figure}
\vspace*{-10mm}
\begin{center}
\mbox{\epsfxsize=16cm\epsffile{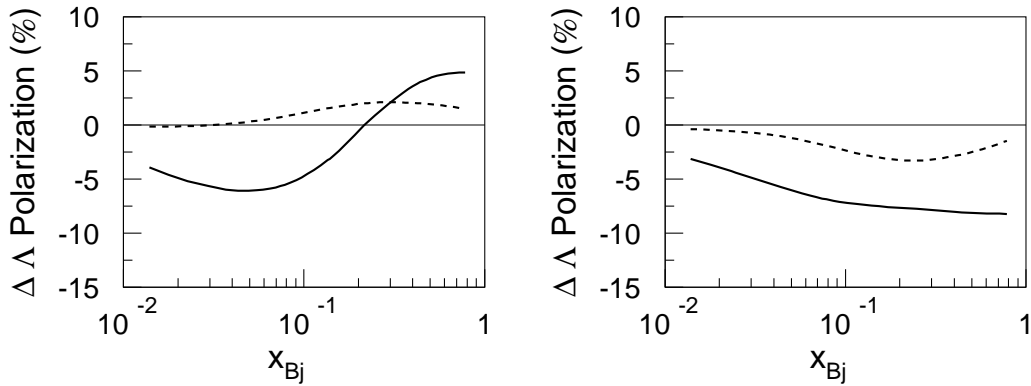}}
\end{center}
\vspace*{-5mm}
\caption{Comparison of $\Delta P_\Lambda$ for two different polarized
parton densities 
(solid line - Brodsky, Burkardt, and Schmidt~\protect\cite{bbs95} and
dashed - Gehrmann and Stirling~\protect\cite{GS94})
for the {\it BGH} (left plot) and the {\it BJ-I} (right plot)
spin transfer mechanism.}
\label{fig:polpdf}
\end{figure}

In Figure~\ref{fig:polarized} we present our predictions for the \la~(\al)
polarization difference $\Delta P_\Lambda$ ($\Delta P_{\overline \Lambda}$).
As input for the polarized quark distribution functions we used the 
Brodsky, Burkardt, and Schmidt parametrization~\cite{bbs95},
which predicts a large negative sea quark polarization
($\Delta {\sf s} = -0.10$, and $\Delta s(x) / s(x) \sim -0.20$ at
$x \sim 0.1$ and input scale of $Q^2_0 = 4~{\rm GeV}^2$).
Different polarized parton densities were also considered
(see Figure~\ref{fig:polpdf}).

All the different spin transfer models lead to similar predictions
for $\Delta P_\Lambda$ ($\Delta P_{\overline \Lambda}$) within a few~\%,
except for \la's produced at high $x$.
This effect is easily understood, given that at high $x$,
\la~production is dominated by scattering off polarized \uq~quarks,
which transfer their polarization to the \la's in different ways
and with opposite signs (see Table~\ref{tab:bgh} and Table~\ref{tab:bjsr}).
The {\it dip} in the \lal~polarization distributions
just below $x \sim 0.1$
originates from the large negative \sq/${\sf \bar s}$ polarization
in this $x$ interval correlated with a large positive spin
transfer coefficient $C_s^\Lambda$,
and the positive \uq/${\sf \bar u}$ polarization with a negative spin
transfer coefficient $C_u^\Lambda$;
therefore both quarks contribute with the same sign.
These results were obtained with the  polarized parton density parametrization
of Brodsky, Burkardt, and Schmidt~\cite{bbs95}.

For comparison we also used different parametrizations of the polarized
quark distributions,
like the Gehrmann and Stirling one~\cite{GS94} with a zero
sea quark polarization at the input scale ($Q^2_0 = 4~{\rm GeV}^2$).
With this parametrization we obtained considerably smaller results.
Near zero values for the polarization difference
$\Delta P_\Lambda$ ($\Delta P_{\overline \Lambda}$)
were obtained using the polarized parton densities of
Gl\"uck, Reya, Stratman, and Vogelsang~\cite{GRSV}.
Table~\ref{tab:polpdf} summarizes the \lal~longitudinal polarization results,
integrated in $z$ for $x_F > 0$ and $z > 0.2$,
using these three different polarized parton densities.
Figure~\ref{fig:polpdf} compares the $\Delta P_\Lambda$
expectations for the first two
polarized parton densities using the {\it BGH} and the {\it BJ-I} 
spin transfer mechanism.

\begin{table}
\begin{center}
\begin{tabular}{|c||c|c|c|c|}
\hline
Pol. part. dens. & $NQM$ & $BGH$ & $BJ-I$ & $BJ-II$ \\ \hline \hline
      $BBS$  &  $-7.2$ & $-4.0$ & $-6.5$ & $-6.0$  \\ \hline
\cite{bbs95} & $-10.3$ & $-7.7$ & $-5.5$ & $-6.5$  \\ \hline \hline
     $GS94$  &   $0.1$ &  $1.1$ & $-2.0$ & $-1.4$  \\ \hline
\cite{GS94}  &   $0.1$ &  $0.1$ &  $0.0$ & $-0.4$  \\ \hline \hline
     $GRSV$  &   $0.0$ &  $0.3$ & $-0.5$ & $-0.3$  \\ \hline
\cite{GRSV}  &   $0.0$ &  $0.0$ &  $0.0$ & $-0.1$  \\ \hline \hline
\end{tabular}
\end{center}
\caption{$\Delta P_\Lambda$ (upper lines) and $\Delta P_{\overline \Lambda}$
(lower lines) for three different polarized parton densities
and different spin transfer mechanisms.}
\label{tab:polpdf}
\end{table}

The longitudinal polarization of \lal's produced in the scattering
off a polarized nucleon, should allow, at least in principle, to access the
polarized quark densities in the nucleon, once the spin transfer mechanism
for the \lal~production is understood.
>From the study with an unpolarized target (previous Section),
the $\Delta D_q^\Lambda$ which best describes the data can be determined,
and then used for the polarized target case.
However, our studies have shown that typically one would expect at most
$|\Delta P_\Lambda| \sim 6~\%$: for a solid target, like in most fixed
target experiments, $|\Delta P_\Lambda|$ reduces to only 1--2~\%.
In addition, one can determine, realistically, the \lal~longitudinal 
polarization with a precision not higher than a percent,
because of experimental systematic uncertainties in the measurement
of this quantity.
These facts indicate a relatively small sensitivity of the
\lal~polarization to the target polarization,
contrary to what is expected for instance in Ref.~\cite{lu}.
Therefore, only a crude estimate of the polarized parton densities
can be obtained through the study of the \lal~polarization.
Nevertheless, a sizeable (negative) \lal~polarization
would indicate a large (negative) {\it strange} sea polarization.

\section{\la~polarization in neutrino and anti-neutrino production}
 
Particularly interesting conditions for the measurement
of polarized fragmentation functions are provided by \la~production
in neutrino and anti-neutrino DIS.
In neutrino scattering the flavor changing charged current weak interaction
selects left-handed quarks (right-handed anti-quarks),
giving 100~\% polarized fragmenting quarks.
For this process Eq.~\ref{eq:lambdaup} reads
\begin{equation}
P_\Lambda \, (x,z) = 
   \frac{\sum_{q,q^\prime} \epsilon_q w_{qq^\prime} \; q(x) \;
          \Delta D_{q^\prime}^{\Lambda}(z)}
     {\sum_{q,q^\prime} w_{qq^\prime} \; q(x) \; D_{q^\prime}^{\Lambda}(z)}.
\label{eq:lambdanu}
\end{equation}
where the $w_{qq^\prime}$ are the $W^+q$ ($W^-q$) weak charge couplings
(for instance $w_{su} = \sin \theta_C$ where $\theta_C$ is the Cabibbo angle),
$\epsilon_q=-1$ ($\epsilon_{\bar{q}}=+1$) for scattering off (anti-)quarks,
$q$ is the struck quark, and $q^\prime$ is the fragmenting quark
(of different flavor).

Using the {\tt LEPTO} event generator we have performed calculations
for the \lal~polarization in neutrino and anti-neutrino DIS in the
current fragmentation region ($x_F > 0$ and $z > 0.2$) in the same fashion 
as for the electro-production DIS case by implementing Eq.~\ref{eq:lambdanu}
into the event generator code.
In our calculations we used a neutrino and an anti-neutrino
beam of $E_{\nu({\overline \nu})} =50~{\rm GeV}$
incident on an isoscalar target.

\begin{figure}
\vspace*{-10mm}
\begin{center}
\mbox{\epsfxsize=16cm\epsffile{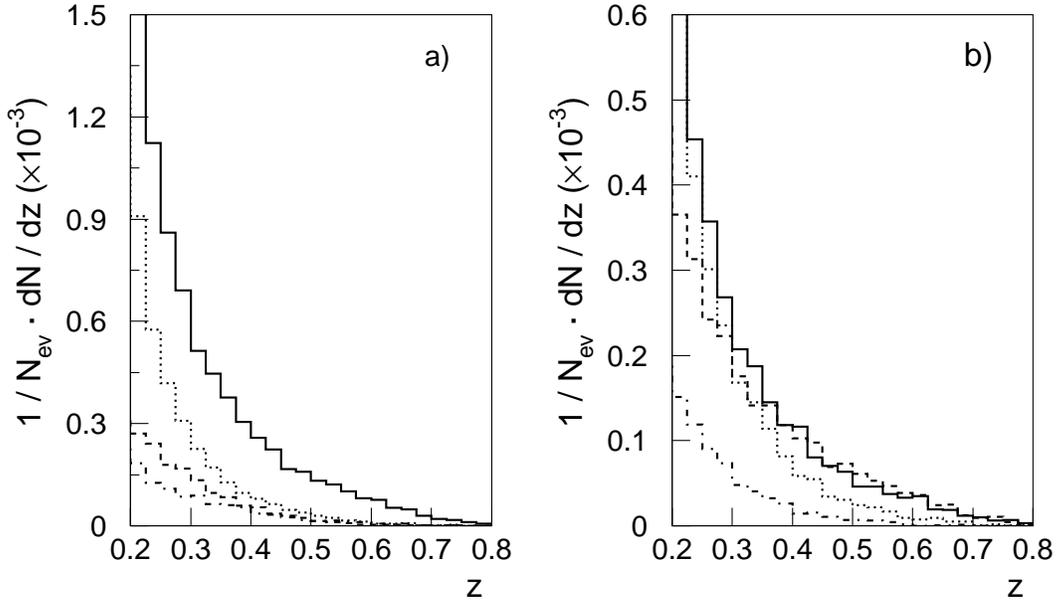}}
\end{center}
\vspace*{-5mm}
\caption{a) normalized $z$ distribution of \la's originating from the
fragmentation of different quark flavors produced in $\nu$--DIS:
solid line - \uq~quark, dashed - \cq, 
and \la's produced in ${\overline \nu}$--DIS:
dotted - \dq, and dot-dashed - \sq;
b) normalized $z$ distribution of directly produced \la's (solid line),
\la's coming from $\Sigma(1385)$ resonances (dotted),
$\Sigma^0$ decays (dashed),
and charmed hadrons (dot-dashed) in $\nu$--DIS.}
\label{fig:nuzed}
\end{figure}

In Figure~\ref{fig:nuzed}a we show the normalized $z$ distribution 
(to the total number of generated events)
of \la's created in the fragmentation of different quark
and anti-quark flavors as obtained from the {\tt LEPTO} code.
Figure~\ref{fig:nuzed}b shows the $z$ distribution of directly produced \la's,
as well as \la's coming from $\Sigma^0$ decays, higher spin resonances
$\Sigma(1385)$, and cascades.
These distributions (Figure~\ref{fig:nuzed}b) are similar to the ones obtained
with a muon beam (see Figure~\ref{fig:muzed}b).

\begin{figure}
\vspace*{-10mm}
\begin{center}
\mbox{\epsfxsize=16cm\epsffile{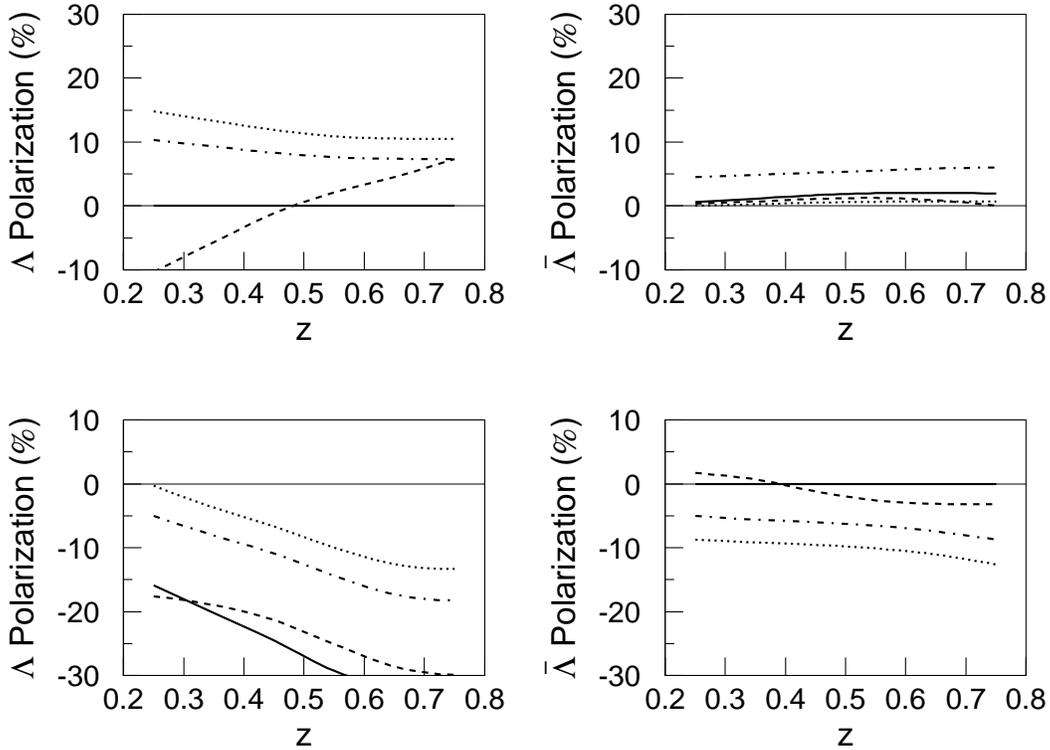}}
\end{center}
\vspace*{-8mm}
\caption{\lal~polarization in the current fragmentation region in 
$\nu$--DIS (upper plots) and $\overline{\nu}$--DIS (lower plots):
solid line - $NQM$, dashed - $BGH$, dotted - {\it BJ-I}, and
dot-dashed - {\it BJ-II}.}
\label{fig:neutrino}
\end{figure}

The results for the \lal~polarization as a function of
$z$ are presented in Figure~\ref{fig:neutrino}.
In neutrino scattering, \la~production is dominated by fully polarized
fragmenting \uq~quarks (see Figure~\ref{fig:nuzed}),
with a small contribution of \cq~quarks ($\sim 10~\%$) at this energy
(at a lower energy the contribution of \cq~quarks is smaller).
In ${\overline \nu}$--DIS both \dq~and \sq~quark fragmentation 
contributes to \la~production
and the latter dominates at large $z$.
Note that the polarization of \dq~and \sq~quarks is opposite
compared to that of \uq~quarks in $\nu$--DIS.
This easily explains the observed behavior of the
\la~polarization for the considered mechanisms of the spin transfer
shown in Figure~\ref{fig:neutrino}.

The difference between the {\it constituent quark} and the $g_1^\Lambda$
{\it sum rule} spin transfer mechanism are quite significant in neutrino
scattering.
These two schemes lead to different signs for the $\Lambda$ polarization,
contrary to what was found in the $\mu$--beam case (Section~3)
and at the $Z^0$ pole (next Section).
This effect is mainly due to the different signs in the spin transfer
from \uq~quarks, which in this reaction dominate the \la~production.
The {\it NQM} gives zero polarization, since no \sq~quarks are involved,
while the {\it BGH} increases with $z$ from negative values to slightly
positive values at large $z$, which is correlated with the relative
abundances of $\Sigma^\ast$ hyperons.
The {\it BJ-I} and {\it BJ-II} prescriptions give almost constant (in $z$) positive values for the \la~polarization.

The longitudinal \la~polarization in anti-neutrino scattering 
has been measured by the {\it WA59} experiment~\cite{neut}
in the current fragmentation region and kinematical conditions similar
to our analysis.
However, the result obtained $P_{\Lambda}=-0.11 \pm 0.45$ has a large
uncertainty and it is inconclusive as far as this analysis is concerned. 
New data on \lal~production with a neutrino beam at a mean beam energy of
30~GeV will be soon available from the {\it NOMAD} experiment~\cite{nomad}.
At 30~GeV we expect similar \la~polarization results as those shown in
Figure~\ref{fig:neutrino}.
This experiment might collect a sample of several thousands \la's,
giving a relatively accurate measurement of the \la~polarization
within a few~\% and thus allowing one to distinguish between the
models considered here and to measure directly the polarized
fragmentation function
$\Delta D_{\sf u}^\Lambda$ for \uq~quarks in {\it clean} conditions.

\section{\lal~polarization at the $Z^0$ pole}

The Standard Model predicts a high degree of longitudinal polarizations
for quarks and anti-quarks produced in $Z^0$ decays: 
$P_s=P_d=-0.91, \;P_u=P_c=-0.67$~\cite{KPT94}.
Thus, reaction (\ref{eq:ee}) is a source of polarized
quarks which can be exploited to investigate the 
spin transfer dynamics in polarized quark fragmentation. 

A large \lal~longitudinal polarization 
($P_{\Lambda}=-0.32  \pm 0.06$ for $z>0.3$)
has been recently reported by the {\it ALEPH} collaboration~\cite{aleph}.
The authors concluded that the measured \lal~longitudinal
polarization is well described by the {\it constituent quark} model 
predictions of Gustafson and H\"akkinen~\cite{gh}.
However, as our study shows, the interpretation of this data 
is not unique.

\begin{figure}
\vspace*{-10mm}
\begin{center}
\mbox{\epsfxsize=16cm\epsffile{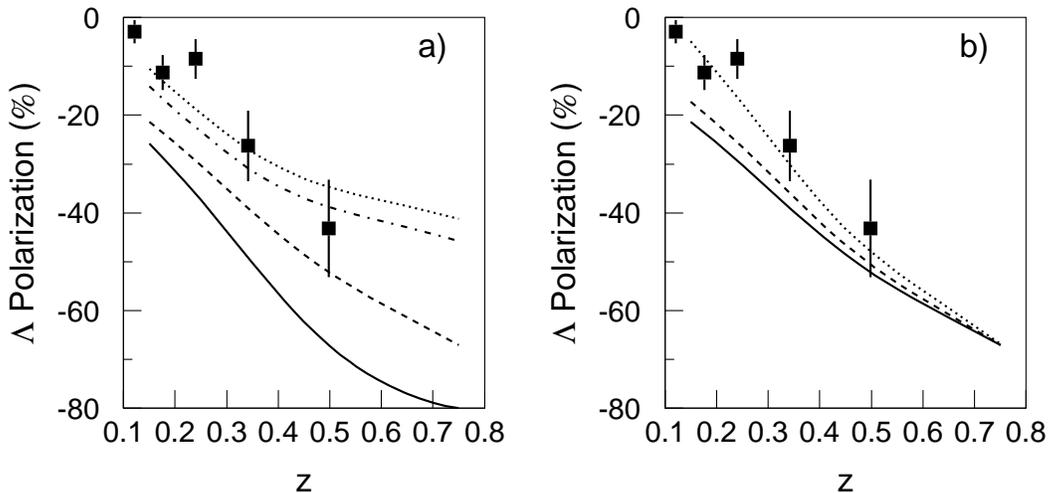}}
\end{center}
\vspace*{-8mm}
\caption{a) \lal~polarization at the $Z^0$ pole for different mechanisms
of spin transfer: solid line - $NQM$, dashed - $BGH$, dotted - {\it BJ-I},
and dot-dashed - {\it BJ-II}.
The experimental data (full squares) are from~\protect\cite{aleph}.
b) comparison between predictions using the {\it BGH} model
for the \la~polarization in our analysis
(solid line) and the analysis of~\protect\cite{aleph} assuming that
only \sq~quarks contribute to \la~polarization (dashed), and additionally
that only first rank \la's inherit a fraction of the fragmenting quark
polarization (dotted).}
\label{fig:lamz0pol}
\end{figure}

In Figure~\ref{fig:lamz0pol}a we present our predictions for different
spin transfer mechanisms for the \lal~polarization at the $Z^0$ pole.
These predictions are compared with experimental data from~\cite{aleph}. 
At high $z$ both models, the {\it BGH} and $g_1^\Lambda$ {\it sum rule},
describe the experimental data fairly well,
while the {\it NQM} mechanism
gives too large values for the \lal~polarization.
At small $z$ the data even favors the $g_1^\Lambda$ {\it sum rule} mechanism.
Also in this case more precise experimental data are needed to distinguish
between the various models of the spin transfer 
mechanism in quark fragmentation.

Our analysis differs from that of~\cite{aleph} in two main points.
The authors of~\cite{aleph} assume in their analysis (similarly as in 
Ref.~\cite{gh}) that only fragmenting polarized \sq~quarks contribute
to the \la~polarization
({\it i.e.} $C_u^\Lambda = C_d^\Lambda = 0$ in Table~\ref{tab:bgh}).
Additionally, they separate between first and lower rank \la's
produced in the string fragmentation, and they assume that lower rank \la's
do not inherit any polarization from the fragmenting quark.
Their argument that \sq~quarks produced in the fragmentation process
have no longitudinal polarization due to parity conservation is not
applicable to polarized quarks fragmentation.
In our study, instead, all quark flavors contribute to
the \la~polarization (according to Table~\ref{tab:bgh}), and
we also do not separate between first and lower rank \la's.
For comparison we also performed the analysis of~\cite{aleph}.
Figure~\ref{fig:lamz0pol}b compares the results thus obtained
for the {\it BGH} spin transfer mechanism.
These two different approaches give similar results in the high $z$ region,
where \la~production is dominated by fragmenting \sq~quarks and
the \la~polarization is directly correlated to that of the fragmenting 
\sq~quark (first rank particle), while at lower $z$ our analysis predicts
larger values for the \la~polarization.

\section{Conclusions}

In this work we have studied several lepton induced processes, where
the longitudinal spin transfer in polarized quark fragmentation can be 
investigated through the measurement of the \lal~longitudinal polarization
produced in these processes.
Two different scenarios for the spin transfer mechanism,
the {\it constituent quark} and the $g_1^\Lambda$ {\it sum rule},
were used for numerical estimates of the \lal~longitudinal polarization.
To distinguish between various spin transfer mechanisms,
it is important to measure the \lal~polarization in different
processes, since different quark flavors are involved in the
fragmentation to a \lal~hyperon with different weights.
For instance, the largest effects are expected in neutrino DIS, where mainly
\uq~quarks fragment to a \la, and the
two scenarios predict also different signs for the polarization.
Typically, the {\it constituent quark} spin transfer mechanism predicts,
in magnitude, larger values for the \lal~polarization, while the 
$g_1^\Lambda$ {\it sum rule} mechanism predicts smaller values.
The existing experimental data have large uncertainties on the
polarization measurements \footnote{ Note that preliminary results
from the {\it DELPHI} collaboration \cite{delphi}
indicates a value for the \la~polarization in the
reaction (2) compatible with zero.} and cannot separate between these models.
Current and future semi inclusive DIS experiments will soon provide
accurate enough data to study these phenomena.

Our studies have shown that the \lal~polarization in
electro-production is less sensitive to the target polarization (in general)
and to $\Delta {\sf s}$ as expected in Ref.~\cite{lu}.
The main physics reason behind this is that \la~production is dominated by
scattering off \uq~quarks even in the low $x$ region.
The production of \la's from scattering off \sq~quarks can be enhanced
by selecting events in the high $z$ region.
However, in this region the \lal~yields drop
significantly, and measurements will be limited by small statistics.
The small sensitivity is also due to the experimental difficulties
in measuring the longitudinal \lal~polarization to very high accuracy and
in the realization of proton targets with a high effective polarization.
Nevertheless, a sizeable (negative) \la~polarization
($i.e.$ $\Delta P_\Lambda$) will indicate a large (negative) polarization
of sea quarks in the polarized nucleon.

\section*{Acknowledgements} 
 
Part of this work was initiated with C.A.~Perez.
We appreciate his contribution and warmly acknowledge his help.
It is also a pleasure to thank our {\it COMPASS} colleagues,
A.~Efremov, J.~Ellis, S.~Gerassimov, and P.~Hoodbhoy for valuable discussions
and comments. We are grateful to L.~Camilleri for discussions on the
{\it NOMAD} experiment, and P.~Hansen for correspondence on the 
\lal~polarization measurement in {\it ALEPH}.



\end{document}